\documentclass[prl,showpacs,twocolumn,amsmath,amssymb]{revtex4}
\usepackage{graphicx}
\usepackage{color}

\def\beq{\begin{equation}}
\def\ee{\end{equation}}
\def\bi{\begin {itemize}}
\def\ei{\end{itemize}}

\def\lsim
{\protect \raisebox{-0.75ex}[-1.5ex]{$\;\stackrel{<}{\sim}\;$}}
\def\gsim
{\protect \raisebox{-0.75ex}[-1.5ex]{$\;\stackrel{>}{\sim}\;$}}
\def\lsimeq
{\protect \raisebox{-0.75ex}[-1.5ex]{$\;\stackrel{<}{\simeq}\;$}}
\def\gsimeq
{\protect \raisebox{-0.75ex}[-1.5ex]{$\;\stackrel{>}{\simeq}\;$}}

\def\beq{\begin{equation}}
\def\ee{\end{equation}}

\def\bi{\begin {itemize}}
\def\ei{\end{itemize}}

\begin{document}

\title{Efficiency of autonomous soft nano-machines at maximum power}
\author{Udo Seifert
}

\affiliation{
{II.} Institut f\"ur Theoretische Physik, Universit\"at Stuttgart,
  70550 Stuttgart, Germany}
\pacs{05.40.-a, 87.16.-b}
\begin{abstract}

We consider nano-sized artificial or biological machines working in 
steady state enforced by
imposing non-equilibrium 
  concentrations
of solutes or by applying external forces, torques or electric fields.
For
 unicyclic and strongly coupled multicyclic machines, efficiency at maximum 
power 
is not bounded by the  linear response value 1/2. For strong
driving, it can  even 
approach the
thermodynamic limit 1. Quite generally, such machines
fall in three different 
classes characterized, respectively, as ``strong and efficient'', ``strong and 
inefficient'',
 and ``balanced''. For weakly coupled  multicyclic machines, 
efficiency at  maximum
power has lost any universality even in the linear response regime.

\end{abstract}

\maketitle

\def\lsim
{\protect \raisebox{-0.75ex}[-1.5ex]{$\;\stackrel{<}{\sim}\;$}}

\def\gsim
{\protect \raisebox{-0.75ex}[-1.5ex]{$\;\stackrel{>}{\sim}\;$}}

\def\lsimeq
{\protect \raisebox{-0.75ex}[-1.5ex]{$\;\stackrel{<}{\simeq}\;$}}

\def\gsimeq
{\protect \raisebox{-0.75ex}[-1.5ex]{$\;\stackrel{>}{\simeq}\;$}}

\def\wo{w_{\rm out}}
\def\bwo{\widehat w_{\rm out}}
\def\wi{w_{\rm in}}
\def\xo{x_{\rm out}}
\def\xi{x_{\rm in}}
\def\etai{\eta^*_{\rm in}}
\def\etao{\eta^*_{\rm out}}
\def\etass{\eta^{**}}

\def\hi{h^{\rm in}}
\def\di{d^{\rm in}_{mn}}
\def\he{h^{\rm out}}
\def\de{d^{\rm out}_{mn}}

\def\Po{P_{\rm out}}
\def\Pii{P_{\rm in}}

{\sl Introduction.--} 
Molecular motors \cite{howard} as well as the recently developed artificial
nano-machines inspired by them  \cite{kay07, bath07}
operate in an aqueous solution
of constant temperature. In contrast to heat engines limited by
Carnot's law, thermodynamics constrains their efficiency by 1.
Like for heat engines, however, operating at the upper bound
comes at the price of zero power since it requires infinitely
slow driving. A
practically more relevant question then is about efficiency at
maximum power (EMP). For heat engines this problem has a few decades 
of history \cite{curz75}
where the full picture exhibiting some universality
close to the linear response regime has just emerged 
\cite{vdb05,schm08,espo09,espo10c}.

For machines working under isothermal conditions, which provide
arguably a much more realistic route for a successful
implementation on the nano to micro scale than dealing with
heat baths of different temperature, efficiency has been discussed
mostly in the context of molecular motors 
\cite{parm99,dere99,wang02a,parr02,kolo07,gasp07,lau07a,liep10}. 
The issue of EMP, however, has received much less
attention so far. For molecular motors, a case study has shown
values well above the linear response result
1/2 \cite{schm08a}. On the other hand, this value has recently
been derived as a universal bound even beyond the linear response regime
under conditions
claimed to be
 relevant
for mesoscopic machines with a sufficiently large number of
internal states \cite{gave10}.

In this paper, we address the issue of EMP for
autonomous soft nano-machines using a quite general approach
requiring minimal assumptions. ``Autonomous'' stands for steady
state conditions as typical for molecular motors and envisaged
for artificial machines. At present, the latter
 mostly still need 
a modulation of external parameters for
driving them through a cycle as investigated theoretically
in Refs. \cite{astu07,raha08,cher08}.
``Soft'' is short for working
in an aqueous environment which will require some flexibility
in the molecule(s) allowing for conformational changes to
operate as a machine. By implementing thermodynamic consistency
from the very beginning and by separating universal from system
specific quantities, the result will be applicable to a large 
class of machines which turn out to fall into three
regimes. In particular, we  find that EMP
 is not bounded by 1/2. It can rather 
approach 1, which, however, requires some care in the 
design of the machine as well as a sufficiently large
parameter space available for the maximization of power.
In fact, for a severly restricted parameter space, EMP 
looses any universality.

{\sl Model.--} We model the kinetics of the nano-machine as a
Markov process in a heat bath of constant temperature  
\cite{schn76,hill,seif10}. 
At any time, the machine is in one of several possible states.
A transition between state $m$ and state $n$ happens
with a rate $k_{mn}$. Typically, in a transition, some (external)
quantities 
$d^\alpha _{mn}$ later
to be associated with the function of the machine change. Examples for such
quantities could be (i) position along a linear track, (ii) rotation angle,
(iii) number of consumed (or, if negative, produced) molecules of  a
certain species, or (iv) number of charges transported against an external
field. 
For each such quantity, there exists a
``conjugate'' external field $h^\alpha$ with the property that the product
$h^\alpha d^\alpha_{mn}$ is a (free)  energy. Specifically, for the above
cases, the conjugate
field is
(i) force $f$, (ii) torque $N$, and (iii) deviation $\Delta \mu^i$  of chemical potential of 
species $i$ from its equilibrium value and (iv) potential
difference $\Delta \phi$.

For the object to operate as a useful machine, at least two of the fields,
an input field $\hi$ and an output field $\he$,
have to be set to non-zero values. For a rotary  molecular motor like
the F1-ATPase (see, e.g. \cite{toya10,haya10} and Refs. therein), one
has to provide ATP at higher than equilibrium chemical potential, i.e.,
 $\hi=
\Delta \mu^{\rm ATP}$. Hydrolysis of ATP can then be used to move against an
external torque, i.e., $\he = N$. In contrast to macroscopic machines,
the role of input and output is not 
unique but can be interchanged depending on the intended ``purpose'' 
of the machine. The ATPase can deliver mechanical work
if fed with an excess of ATP molecules but can also synthesize ATP if pulled
by an external torque in the opposite direction.

The external fields will affect the  transition rates. Thermodynamic consistency requires a
local detail balance (LDB)  condition of the form 
\beq
\frac{k_{mn}(\hi,\he)}{k_{nm}(\hi,\he)} =\left(\frac{k_{mn}}{k_{nm}}
\right)_{\rm eq}
\exp(\hi\di+\he\de)
\label{eq:ldb}
\ee 
where $(k_{mn}/k_{nm})_{\rm eq}$ refers to the equilibrium ratio where all
external fields are set to zero.
Here, and throughout the paper, we measure energies in thermal units ($k_BT=1)$.
 For constant external fields, $\hi$ and $\he$, the machine reaches
a steady state, i.e., will operate autonomously.

{\sl Unicyclic machines.--}
We first discuss unicyclic machines, for which the $m=1, ..., M$ possible 
states are
aligned in one cycle such that each state has two neighboring states. 
If the machine
steps through the complete cycle in forward direction, it has consumed
the work or (free) energy
$
\wi = \sum_{m=1}^M \hi d^{\rm in}_{m,m+1}
$ and delivered the output 
$
\wo \equiv -  \sum_{m=1}^N\he  d^{\rm out}_{m,m+1} 
$
where $ d^{\rm in}_{M,M+1}\equiv  d^{\rm in}_{M,1}$.
 Likewise, had it stepped through the complete cycle in backward 
direction, it would have released $-\wi$ to the input reservoir
and consumed $-\wo$ from the output reservoir. In order to work as a machine
in the conceived sense, the mean  time $\tau^+$ it takes to complete the
cycle in forward direction has to
be smaller than the mean time $\tau^-$ for completing the cycle backwards. 
Both times can
be expressed diagrammatically in terms of the transition rates 
but the explicit expressions  will not
be needed here \cite{hill}. The key point is that the ratio $\tau^+/\tau^-$ is given
by the ratio between the product of all backward rates and the product of
all forward rates. With the local detail balance condition,
eq. (\ref{eq:ldb}),
and the identifications 
of input and output work just given,
 this ratio becomes
\beq
\tau^+/\tau^- = e^{\wo-\wi} .
\label{eq:tautau}
\ee
Hence the power (or, more generally, the rate of 
``yield'' if the output
are chemical products) 
delivered by the  machine in the steady state is
\beq
\Po = \wo(1/\tau^+-1/\tau^-)=\wo[1-e^{\wo-\wi}]/\tau^+ .
\label{eq:pow}
\ee Likewise, the power used by the motor becomes
\beq
\Pii = \wi(1/\tau^+-1/\tau^-)=\wi[1- e^{\wo-\wi}]/\tau^+ .
\ee
 These transparent expressions
for the power
where the specific characteristics of the motor
enter through the single quantity $\tau^+$ constitute our first main
result. In the regime $0<\wo<\wi$,
the machine will work as intended. Its efficiency is simply
given by
$
\eta \equiv \Po/\Pii=\wo/\wi
$ and is obviously bounded by thermodynamics through $0 < \eta < 1$.
For $\wo=\wi$, the machine has optimal efficiency $\eta = 1$ 
but does not deliver
any power  since it then cycles as often in forward as in backward
direction.

The concept of EMP requires to  identify
 the admissible variational parameters. Rather than starting with an
arbitrary parameter set  $\{\lambda_i\}$, the form 
of eq. ({\ref{eq:pow}) strongly
suggest to focus first on the following three choices:
(i) $\wo$, (ii) $\wi$, and (iii) $\wo$ and $\wi$. For each case,
the crucial question becomes how the forward cycle time $\tau^+$ depends on
input and output. We will first assume that
\beq
\xo\equiv - d\ln \tau^+/d\wo {\rm ~~~ and ~~~}\xi\equiv - d\ln \tau^+/d\wi
\nonumber
\ee
are constants, i.e., that $\tau^+$ depends mono-exponentially (with either
sign) on input and output work as illustrated below with a specific
example and later discuss the general case.

{\sl Maximization with respect to output $\wo$.--}
For given input $\wi$, the condition $d\Po/d\wo=0$ leads to the 
implicit relation 
\beq
\wi = \wo^* + \ln \left(1+ \frac{\wo^*}{1+\xo\wo^*}\right) 
\label{eq:win}
\ee
for the optimal 
output $\wo^*$
at fixed input $\wi$. Consequently, we obtain for EMP
under these conditions
\beq
\etao\equiv \frac{\wo^*}{\wi} = \left[ 1+
  \frac{1}{\wo^*}\ln\left(1+\frac{\wo^*}{1+\xo\wo^*}\right)\right]^{-1} .
\label{eq:etao}\ee
 Two limit cases can be discussed analytically.

First, for small deviations from thermodynamic equilibrium, $\wi\ll1$, expanding 
(\ref{eq:win}) and (\ref{eq:etao})
leads to $\wo^*\approx \wi/2$ and
to an efficiency at maximum
power given by 
\beq
\etao = 1/2 + (\xo+1/2)\wi/8 + O(\wi^2).
\nonumber
\ee 
 For $\xo > -1/2$, 
EMP increases if the machine operates beyond the
linear response regime thus beating the
bound 1/2 found in Ref. \cite{gave10} under different conditions.

Second, for large $\wi\gg1$,  two cases
must be distinguished. If $\xo > 0$, EMP,
$
\etao \approx 1-[\ln(1+1/\xo)]/\wi
$,  approaches the thermodynamic 
upper bound 1. If $\xo<0$, however, $\wo\to 1/|\xo|$ for 
$\wi\to \infty$
which implies 
$
\etao\approx 1/|\xo \wi| .
$
In this case, efficiency at optimal power approaches 0 with increasing input
since maximum power is reached for a finite $\wo^*$.

{\sl Maximization with respect to input $\wi$.--}
We now assume that the output $\wo$ is given and search for the optimal
input $\wi^*(\wo)>\wo$ maximizing the power. Two cases must be distinguished.
(i) For $\xi>0$, the power grows unbounded 
as $\wi$ increases and hence EMP 
vanishes for all $\wo$.
(ii) For $\xi<0$, the optimal input reads
$
\wi^*=\wo+\ln(1+1/|\xi|)
\label{eq:wii}
$
implying the EMP\beq
\etai = 
\frac{\wo}{\wi^*} = \left[ 1+
  \frac{1}{\wo}\ln\left(1+1/|\xi|\right)\right]^{-1} .
\nonumber
\ee
For $\wo\ll1$, the efficiency
$
\etai\approx \wo /\ln\left(1+1/|\xi|\right) 
$
differs from the linear response result derived above for
output maximization since the optimal $\wi^*$ 
is always a finite distance from
$\wo$ and hence no longer within the linear response regime.
For $\wo\gg 1$, one gets the asymptotics
$
\etai\approx 1-[\ln(1+1/|\xi|)]\wo.
$

\begin{figure}
\includegraphics[width=8cm]{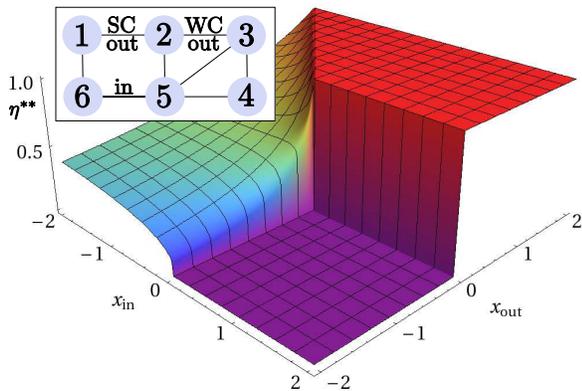}
\caption{Efficiency at maximum power $\eta^{**}(\xi,\xo)$.
Inset: Example of a
multicyclic machine with 6 states, 6 cycles and the input transition (56). If (12) is the
output transition (or, equivalently, (56) or (12)) the machine is 
strongly coupled (SC).
For (23) as  output transition, it is weakly coupled (WC). In the latter case, $C_1$
consists of cycles (123456) and (12356), $C_2=\{(2345),(235)\}$, 
$C_3=\{(1256)\}$, and $C_4=\{(345)\}$.}
\label{fig1}
\end{figure}

{\sl Maximization with respect to input and output.--}
This case, solved by $\wo^{**}$ and $\wi^{**}$, leads to three
regimes for EMP, 
$\etass\equiv \wi^{**}/\wo^{**}$, shown in Fig. \ref{fig1}.

(i) For $\xo>{\rm max}\{-\xi, 0\}$, the power at optimal output increases
exponentially with growing input. In this case, one has $\wi^{**}\to \infty$
and $\wo^{**}\to \infty$ with the finite difference
$
\wi^{**}-\wo^{**} \approx \ln(1+1/\xo) 
$
and hence $\etass = 1$. Such machines could be called ``strong and
efficient''.

(ii) For $\xo<0$ and $\xi>0$, the power still increases with increasing input,
and, hence $\wi^{**}\to \infty$.
The optimal output, however, in this limit remains finite at
$\wo^{**}=1/|\xo|$ which implies $\etass = 0$ for these
``strong but inefficient'' machines.

(iii) Finally, for $\xi<{\rm min}\{-\xo,0\}$, the power peaks at a genuine
maximum for $\wo^{**}=1/[-(\xi+\xo)]$ and  
$\wi^{**}=\wo^{**}+\ln(1+1/|\xi|)$ leading to 
\beq
\etass=\wo^{**}/\wi^{**}=1/\left[1+|\xo+\xi|\ln(1+1/|\xi|)\right] .
\nonumber
\ee Such machines could be called ``balanced''.

{\sl Example.--}
For a simple but still instructive specific example, we consider a
one-state molecular motor
 for which hydrolysis of ATP leads to a forward step of
length $d$ along a linear track in the direction of an applied force
$f$ with rate 
$
k^+=1/\tau^+ = k_0 \exp [\Delta \mu^{\rm ATP} + f\theta^+ d] ,
$ where $k_0$ is the equilibrium rate. A backward step with rate 
and
$
k^-=1/\tau^-=k_0\exp [\Delta \mu^{\rm ADP} + \Delta \mu^{\rm P_i} - f\theta^- d] ,
$ involves synthesis of ATP from ADP and P${\rm_i}$.
 The load sharing factors $\theta^+$ and $\theta^-$ with $\theta^++\theta^-=1 $ 
guaranteeing the LDB condition eq. (\ref{eq:ldb}) are
related to the distance of the activation barrier  in forward and
backward direction, respectively \cite{kolo07}. 
The molecular architecture thus determines the sign of 
$\xo=\partial_{\wo}\ln k^+=-\theta^+$ since $\wo=-fd$.

The input $\wi = \Delta \mu^{\rm ATP} -\Delta \mu^{\rm ADP} - \Delta \mu^{\rm
  P_i}$ 
can be changed in a variety of ways. The most obvious one,
increasing $\Delta \mu^{\rm ATP} $, leads to $\xi= 1$, whereas decreasing 
$\Delta \mu^{\rm ADP}$ leads to $\xi=0$. If both chemical potentials are
changed simultaneously, any value for $\xi$ can be reached, in principle.
Thus, depending on the molecular architecture, in the
EMP diagram, Fig. \ref{fig1},
this motor can either
cross from balanced to strong and efficient (for $\theta^+<0$)
 or from balanced to strong and inefficient (for $\theta^+>0$)
as the way the input power is delivered is changed.

{\sl Arbitrary unicyclic machines.--}
So far, the complete analysis benefitted from using two simplifying assumptions
which will not always apply. First, we assumed the dependence of the
forward cycle time $\tau^+$ to be mono-exponential in both input and output.
While each
individual rate will typically depend mono-exponentially on $\wo$ and $\wi$, 
especially for intermediate values of $\wo$ and $\wi$ several rates may
contribute comparably to $\tau^+$. Then $\xo$ becomes a function
$\xo(\wo,\wi)$ where, e.g., for maximization
with respect to output, $\wo$ has to be determined self-consistently
from eq. 
(\ref{eq:win}). 

Correspondingly, in the analysis of the limiting values, the
quantities $\xo$ and $\xi$ have then to replaced by their respective asymptotic 
limits. 
Generically, in a unicyclic machine consisting of several states,
increasing the input will speed up  one (or several) forward transitions.
If at least one forward transition is not affected by
the increasing input, it will act as a bottleneck for the whole cycle
and, hence, $\xi\approx 0$ for large $\wi$. 
On the other hand, increasing the load,
i.e., increasing $\wo$ will typically slow down at least one forward transition
now becoming the bottleneck 
and consequently $\xo <0$. Thus, generically, according to the EMP diagram, Fig. \ref{fig1}, such machines at maximum power
will approach zero efficiency. In practice, increasing $\wi$ will not
lead to a significantly larger power since $\tau^+$ will approach a limit
value for large $\wi$. At any finite $\wi$, one may thus still reach a reasonable
efficiency as given by eqs. (\ref{eq:win}) and (\ref{eq:etao}) and miss 
maximum power 
by only a small amount.

The second major restriction so far was to focus on $\wi$ and $\wo$ as
control parameters. For full generality,  one should consider a set
of arbitrary  parameters $\{\lambda_i\}$. If the admissible range of these
parameters spans the whole sector $0<\wo<\wi<\infty$ (which will require
at least two control parameters) then we are back at the case just discussed.
On the other hand, if one has only one control parameter, or if only a limited
part of the above sector can be coverered, EMP will become
a strongly  non-universal concept and its specific value will depend on the
functional dependence of $\wo, \wi$ and $\tau^+$ on $\{\lambda_i\}$.

{\sl Multicyclic machines.--}
In general, a nano-machine will contain several cycles. We will assume
that the input affects only one transition, the ``input transition'' and  that there
is only one ``output transition'' which may or may not be the same as the input 
transition.  
Two generic cases, strong coupling (SC) and weak coupling (WC)
 should then be distinguished, see inset of Fig. \ref{fig1}. 

(SC): If the input and the
output transition are identical, or if there exists a direct connection between the two
without any
intermediate bifurcation, any cycle containing the input transition will also
contain the output transition. For such strongly coupled machines exactly the
same formalism as for unicyclic machines applies with the only caveat
that $\tau^+$ appearing there is now given by 
$
1/\tau^+\equiv  \sum_i1/\tau^+_i  
$ where the sum runs over all cycles that include input and output transition
and the $\tau^+_i$ are the corresponding forward cycle times.
Thus, such strongly coupled machines obey the same relations for efficiency
and
EMP as discussed above for unicyclic machines. Note,
however,  that as the explicit expressions would show 
the $\tau^+_i$ are affected by the rates contributing to the cycles not
containing
the input and output transition.

(WC): If the strong-coupling condition is not fulfilled, each 
cycle of the machine
falls  into one of four disjoint subsets $C_1,C_2,C_3,C_4$ containing,
respectively, 
input and output transition, only the output transition, only the input transition, and 
neither one. Since only $C_1$ and $C_2$ contribute to the
output,  the  power is  given
by
\begin{eqnarray}
\Po &=& \wo[(1/\tau^+_1-1/\tau_1^-) + (1/\tau^+_2-1/\tau^-_2)] \nonumber\\
 &=& \wo[(1-  e^{\wo-\wi})/\tau_1^+-(e^{\wo}-1)/\tau^+_2] ,\nonumber
\end{eqnarray}
 where 
$
\tau^\pm_{1,2}\equiv \sum_{i\in {C_{1,2}}} 1/\tau^\pm_i .
$ 
For the second step, we have adapted the local detail balance constraint,
eq. (\ref{eq:ldb}), and its consequence eq. (\ref{eq:tautau})
keeping in mind that  the corresponding ratio for the $C_2$
cycles involves only the output transition. In fact, the presence of these
 cycle necessarily decreases the power.
Moreover, 
stall conditions ($\Po=0$) are now reached at a maximum
value 
$\bwo$
strictly smaller than $\wi$. 
For the input power, one obtains similarly
$
\Pii = \wi[(1- e^{\wo-\wi})/\tau_1^++(1-e^{-\wi})/\tau^+_3] , 
$
where $\tau^+_3\equiv \sum_{i\in {C_{3}}} 1/\tau^+_i.$
These expressions show that both, efficiency $\eta\equiv \Po/\Pii$ and
 EMP $\etao$,  are even less
universal than for strongly coupled machines. For the latter quantity,
this fact becomes obvious by looking at the linear
response regime. Expanding the above expressions for small $0<\wo<\wi\ll1$,
one finds 
\beq
\etao = \frac{1/2}{1+2(\tau_1^+/\tau^+_2 +\tau_1^+/\tau^+_3 
+\tau_1^{+2}/\tau^+_2\tau^+_3)} + O(\wi),
\nonumber
\ee with all cycle times evaluated in equilibrium. Thus, $\etao$ is strictly
smaller than 1/2. Still, one should not conclude from this result that
$\etao$ necessarily remains bounded by 1/2 with increasing $\wi$.

{\sl Concluding perspective.--}
For soft machines working under isothermal steady state
conditions, we have investigated EMP in the appropriate 
parameter space by varying output
power, input power and both. For the first two cases, EMP
is given by a one parameter function whereas in the
third case machines universally fall into three classes.
Our results hold for both unicyclic machines and 
strongly coupled multicyclic ones. Weakly coupled multicyclic
machines are less efficient and their 
EMP is less universal.  Our analytical  results
and the suggested classification provide not only
a transparent theoretical framework but, in a longer term 
perspective, should also be helpful for designing efficient machines. 
From a theoretical point of view, as a next step, machines
operating under periodically modulated fields
should be analyzed similarly for EMP 
since the extant artificial machines typically 
still require 
such conditions.

\end{document}